\documentclass[12pt]{spieman}
\usepackage{amsmath,amsfonts,amssymb}
\usepackage{graphicx}
\usepackage{setspace}
\usepackage{tocloft}
\usepackage[english]{babel}
\usepackage[utf8x]{inputenc}

\usepackage{lastpage}
\usepackage[colorinlistoftodos]{todonotes}
\usepackage{url}
\usepackage{fancyhdr} 

\usepackage{geometry} 
\usepackage[square,sort,comma,numbers]{natbib}	
\usepackage[export]{adjustbox}
\usepackage{booktabs} 
\usepackage{caption}
\usepackage{subcaption}
\usepackage{color}
\usepackage{xcolor,colortbl}
\usepackage{refstyle}
\usepackage{wrapfig}
\usepackage{enumitem}
\usepackage{selinput}

\usepackage{multicol}

\usepackage{framed}
\usepackage{array,ragged2e} 
\usepackage{epstopdf}
\usepackage{titlesec}
\usepackage[toc,page]{appendix}
\usepackage[normalem]{ulem}

\definecolor{forestgreen(web)}{rgb}{0.13, 0.55, 0.13}
\definecolor{applegreen}{rgb}{0.55, 0.71, 0.0}

\definecolor{lightgray}{rgb}{0.83, 0.83, 0.83}
\usepackage{longtable,tabularx,ltxtable}

\title{Genetic Algorithms for Lens Design: a Review}

\author[a]{Kaspar H{\"o}schel}
\author[b]{Vasudevan Lakshminarayanan}
\affil[a]{TU Wien, Karlsplatz 13, 1040 Wien, Austria}
\affil[b]{University of Waterloo, Theoretical and Experimental Epistemology  Lab, School of Optometry \& Vision Science, Depts. of Physics, ECE and Systems Design Engineering, Waterloo, ON, N2L 3G1, Canada} 

\cftpagenumbersoff{figure}
\cftpagenumbersoff{table} 

\newcommand{\tabitem}{~~\llap{\textbullet}~~}
\usepackage{array}
\usepackage{acronym}
\newcolumntype{L}[1]{>{\raggedright\arraybackslash}a{#1}}
\newcolumntype{C}[1]{>{\centering\arraybackslash}p{#1}}
\newcolumntype{R}[1]{>{\raggedleft\arraybackslash}p{#1}}

\hypersetup{
    colorlinks=true, 
    linkcolor=blue,
    filecolor=magenta,      
    urlcolor=cyan,
}
\usepackage{lmodern}

\begin{document} 

\maketitle


\vspace{-0.8cm}
\begin{center}
\line(1,0){430}
\end{center}
\vspace{-0.6cm}

\begin{abstract}
Genetic algorithms (GAs) have a long history of over four decades. GAs are adaptive heuristic search algorithms that provide solutions for optimization and search problems. The GA derives expression from the biological terminology of natural selection, crossover, and mutation. In fact, GAs simulate the processes of natural evolution. Due to their unique simplicity, GAs are applied to the search space to find optimal solutions to various problems in science and engineering.

Using GAs for lens design was investigated mostly in the 1990s, but were not fully exploited. But in the past few years there have been a number of newer studies exploring the application of GAs or hybrid GAs in optical design. In this paper we discuss the basic ideas behind GAs and demonstrate their application in optical lens design.
\end{abstract}

\\
{\noindent \footnotesize{\textsuperscript{a}}E-mail: \linkable{kaspar.hoeschel@student.tuwien.ac.at}}

\vspace{-0.4cm}
\begin{center}
\line(1,0){430}		
\end{center}

\begin{spacing}{1.5}   

\section{Introduction} 
\label{sect:intro} 
In terms of designing optical lenses there are many constraints and requirements, including restrictions like assembly, potential cost, manufacturing, procurement, and personal decision making \cite{thibault2005evolutionary}. Typical parameters include 
surface profile types such as spherical, aspheric, diffractive, or holographic. Usually, the design space for optical systems consists of multi-dimensional parameter space.
Moreover, the radius of curvature, distance to the next surface, material type and optionally tilt, and decenter are necessary for lens design
\cite{Cram2016optometry}.

The most important aspects for designing optical lenses are optical performance or image quality, manufacturing, and environmental requisitions.
\textit{Optical performance} is determined by encircled energy, the modulation transfer function (MTF), ghost reflection control, pupil performance, and the Strehl ratio \cite{edmundoptics2018Intro,fischer2008optical}.
\textit{Manufacturing requirements} include weight, available types of materials, static volume, dynamic volume, center of gravity, and configuration requirements.
Furthermore, \textit{environmental requirements} encompass electromagnetic shielding, pressure, vibration, and temperature. Yilmaz et. al. \cite{yilmazdesign} used a reference temperature of 20°C for designing a lens system. Additional constraints comprise of lens element center and edge thickness, and minimum and maximum air-spaces between lenses. Other important design constraints are maximum constraints on entrance and exit angles, the physically realizable glass index of refraction and dispersion properties \cite{fischer2008optical}. 

Optical designers manufacture a lens system, with all the design requirements for optical lenses, in one place. The most important part of lens design is called optimization. In the process of optimization the values of independent variables (e.g. material between surfaces) are used to realize dependent variables such as imaging magnification \cite{yabe2018optimization}. Furthermore, the optimization process contains local and global minima. Global optimization is needed to find the most stable solution. 
%
%

Traditionally, the most effective optimization tool is the Levenberg–Marquardt algorithm or Damped Least Squares (DLS) method which solves non-linear least squares problems.
%
A disadvantage of DLS is that the designer has to tackle with the local minimum. Therefore global optimization tools were introduced \cite{thibault2005evolutionary,brixner1981lens}. Usually, an optical engineer starts from a global search algorithm with a rough initial configuration of lens design (initial glass selection, number of surfaces, field of view, a wave length and an exact stop position). An automatic optimization algorithm is applied to alter the configuration and to find the best or adequate solution. An initial configuration could be parallel plates of glass to control the surface curvature.

By defining many variables and a merit function the global search algorithms, combined with computer-aided design (CAD) tools, can find design forms with inadequate and many possible solutions.
The optical designer needs to examine one or more optical systems by improving the adjustments and optimizations with algorithms like the \textit{Hammer algorithm} \cite{zemax2011manual} or by hand-calculation. In contrast to traditional optimization tools, the optimization problem of lens design can be solved by the use of a genetic algorithm (GA). This specific kind of algorithm is capable of imitating the principles of biological evolution. A GA is based on repeating the modification of an individual population similar to biological reproduction. Its random nature is utilised to improve the search for a global solution \cite{bajpai2010genetic}.

Since 2015, several researchers have applied the hybrid-GA to lens design \cite{yen2015aspherical1, yen2015aspherical2, tsai2015improvement}. 
These kinds of GAs can be effectively applied to real-world problems, and contain other techniques within their frameworks. It can be argued that hybrid-GAs are more efficient than other types of optimization strategies in the field of lens design. Studies have shown that hybrid GAs can be useful applied for correcting and eliminating chromatic aberrations.


The purpose of this review is to provide an overview about GAs used in lens design. Starting with the background, the optimization problem, and specific requirements of GAs. We then discuss the use of GAs in lens design.\\

\section{Background} 	
\subsection{Optical Lens System}
A search space for lens design encompasses a multidimensional space including several peaks, non-linearity, and a strong correlation between parameters \cite{sturlesi1991future}. The search for local minima is dependent on the initial point solution. Only adjacent points of the initial solution are investigated \cite{thibault2005evolutionary}. Hence, diverse applications of global search methods can be inserted in optical design. Optical software includes special algorithms to investigate beyond optima \cite{forbes1991towards,hearn1991practical,isshiki1998global}.

A typical two-element air spaced lens with nine variables would consist of 4 radii of curvature, 2 glass types, 2 thicknesses, and 1 airspace thickness. Apart from that, a multi-configuration lens includes corrections over the field of view and over a wide spectral band as well as over realistic temperature ranges and over a range of focal lengths. This kind of configuration indicates a complex design volume with many dimensions 
\citep{fischer2008optical}.

Predetermined constraints and parameters are necessary to create an optical lens design. Parameters would include the curvature of spherical surfaces, type of material, and element position. In addition, constraints include magnification, numerical aperture, and field of view. Economic factors incorporate cost, size, and the weight of the system elements. Moreover, the image quality depends on aberrations. The lower the aberrations, the better the image quality and the better the optical lens system   
\cite{hesammahmoudinezhad2014optical}.

\subsection{Lens Optimization}
Typically, there are independent and dependent variables in lens design. Examples for independent variables are such as total surface number, material between surfaces, or the curvature of the surface and dependent variables such as effective focal length, back focal length, or distance from the object surface to the image surface. The most crucial part of lens design is the process of optimization. Optimization deals with receiving independent variables to discover the target values of dependent variables. If the amount of dependent variables is bigger than the amount of independent variables it is not possible to achieve the target values of the dependent variables at the same time, and the problem becomes a least-squares problem \cite{yabe2018optimization}.

\subsection{Problem Areas in Optimization}
Optimization problems have to find the minimum solution of any dimensional problem (e.g. MinMax algorithm, least squares estimation). In the process of optimization, the global minimum or maximum solution is estimated. The global extremum is defined as a point where the function value is smaller or larger than at any other point in the search space.

The local minimum in optimization returns a function value which is smaller than at nearby points in the search space. This value should be greater than at a distant point in the search space \cite{The2018Global}. It is necessary to find as many local minima as possible, because the merit function does not always precisely determine the lens quality. Hence, the designer should select the optimal solution among the local minima \cite{yabe2018optimization}. Typically, finding a global optimum within a search space of many local optima is a challenging problem for all systems which adapt and learn. A genetic algorithm can be used with the right set-up to overcome this deficiency \cite{bajpai2010genetic}. It is the task for the designer to choose a feasible design for an optical system and perform appropriate refinements through numerical modeling \cite{fischer2008optical}. Moreover, the designer is in charge of fulfilling all necessary requirements and adjustments of the optimized lens design or must to restart the entire process again \cite{fischer2008optical}. 

Since the 1940s Baker \cite{Pearce2018J}, Feder \cite{feder1963automatic}, Wynne and Wormell \cite{wynne1963lens}, and Grey \cite{grey1978inclusion} have investigated lens optimization techniques which can overcome this problem in multi-dimensional space. Before the era of digital computers, lens design was calculated manually by the use of trigonometry and logarithmic tables to obtain 2D cuts through multi-dimensional spaces. Computerized raytracing was introduced to facilitate quick lens modeling and the search of design space.

\subsection{Optimization Strategies}
Usually, in nonlinear optimization problems the state-of-the-art technique is called the \textit{Broyden–Fletcher–Goldfarb–Shanno} algorithm \cite{wright1999numerical}. This iterative method is only feasible with available derivatives. Methods like DLS, Newton methods, and variants are common algorithms \cite{kramer2016machine}. Generally, a Quasi-Newton method is required to find the local minima and maxima of functions. 

A GA, or metaheuristic, belongs to a class of evolutionary algorithms (EAs) \cite{jones1998genetic} and is used to simulate and solve optimization problems by applying a population of solutions \cite{chen1997experiment,jakob1999optimierung}. In other words, the GA solves a problem which is encoded in a series of bit strings that are manipulated by the algorithm \cite{Frontline2018genetic}. Furthermore, a GA copes with hard non-linear, multi-modal functions, as well as multi-objective optimization \cite{thibault2005evolutionary}. 

\subsection{Genetic Algorithms}
GAs are both optimization algorithms and heuristic search methods for populations. They are inspired by natural processes, and in particular natural selection and genetic evolution. 

Charles Darwin introduced the terminology of natural selection for the first time in his book \textit{On the Origin of Species} in 1859 \cite{charles1859origin}. Every living organism is related and has a common ancestor. The theory assumes that complex creatures have gradually descended from oversimplified ancestors. Certain random mutation processes occur in the genetic organism's code and particular mutations are kept alive to aid in survival, with follow on generations receiving the last preserved mutations. With time, advantageous mutations will increase and cumulate to produce an entire new generation.

The term genetic algorithm was inspired by Darwin's theory of evolution and first devised by Holland and Goldberg \cite{holland1975adaptation,goldberg1988genetic}. It shows similarities to Rechenberg's \cite{rechenberg1973evolutionsstrategie} \textit{Evolutionsstrategien} (evolutionary strategies ESs) of 1973. Holland's and Goldberg's first approach to this kind of algorithm was theoretical. They used a binary code to describe individuals in a population. This work was later improved by Rechenberg \cite{Adam2004Genetisch}.


GAs are applied to practical problems to assess the solution for a desired outcome, but to also improve the best solution. Examples for practical problems are image processing, prediction protein structures with three dimensions, or in very-large-scale integration. By applying a GA, instead of a specific solution to a problem, the characteristics of the solution are well known. In addition, restrictions of the solutions are used to reject possible and potential solutions. GAs are mostly applied in the field with many large complex problems where conventional algorithms cannot succeed. 
For example the GA can be applied in combinatorial optimization, or parameter estimation 
%
\cite{coley1999introduction}.

Ordinarily, a population of individuals is preserved in a certain search space for a GA. The population represents a possible solution for a given specified problem. Each individual is coded with a finite length vector of variables in the binary alphabet $\left\{{0,1}\right\}$. In the genetic analogy the individuals would resemble chromosomes and variables can be compared to genes. Chromosomes consist of various genes or variables and each chromosome is composed of a binary string. Each bit in the string is characteristic of the solution \cite{bajpai2010genetic}.

\begin{figure}[h]
	\centering
    \caption{Procedure of a GA, modified from \cite{fang2011study}}
    \includegraphics[width=0.5\textwidth]{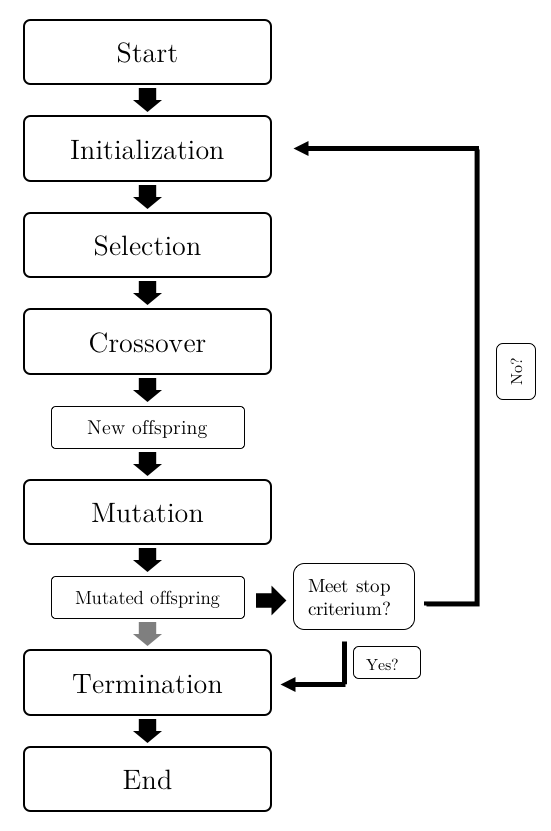}
    \label{fig:procedure}
\end{figure}

When a GA is applied to find a solution in very large problems, it looks into millions of samples from the search space and creates small changes after recombining the best parts of the solution. Then the resultant fitness value is compared with the current best solution and the best solution is taken. The entire process is iterative until a stop condition is met (Fig. \ref{fig:procedure}).\\


There are several benefits of using a GA over other optimization strategies. An important aspect is the use of optimization with a systematized set of continuous or discrete parameters for global optimization scenarios. It works with a large number of parameters. The result of applying a GA gives a set of solutions and not a single solution \cite{cuevas2006genetic}.
On the other hand, high computational cost is required to look into billions of solutions which impede the optimization of hard problems. Fine tuning of all parameters for the GA is associated with trials and errors
\cite{thibault2005evolutionary,whigham2017mapping}. 

\subsubsection{Procedure of a GA}
Generally, a basic genetic algorithm includes five phases: 1. initialization, 2. selection, 3. crossover, 4. mutation, and 5. termination \cite{sheppard2016}.

\begin{enumerate}
\item \textbf{Initialization}\\
A gene set is built out of a population of candidate solutions.
The algorithm generates random strings from the individual solutions to form an initial population. 
The initialization is done randomly to cover the total range of possible solutions in the search space. Normally, the population size is not depending on the nature of the problem, but has a reasonable size of about 100 to 1000 optimal solutions.

\item \textbf{Selection and fitness value}\\
Individual genomes are selected from an existing population to breed the new generation. Individual solutions are selected through a fitness function within a fitness-based process and are named fitter solutions. Selection methods are chosen to rate the fitness of each solution. 
The fitness value is generated to give feedback so the GA can find the best solution. 
Most fitness functions are stochastic methods in order to select small proportions of less fitness solutions and maintain a variety of large populations. Furthermore, these functions prohibit premature convergence on poor solutions. Additional selection methods include roulette and tournament wheel selection.

\tabitem\textbf{Roulette Wheel Selection }
The fitness level is the requirement that each individual solution is linked with a probability of selection. In this method, the fitness value for each input is calculated and depicted on the wheel in portions of percentage (Fig. \ref{fig:roulette}).
The wheel is rotated and has a search space of n-chromosomes. A chromosome with a high fitness value will be selected more than once.

 \tabitem\textbf{Tournament Wheel Selection}\\
This method takes two solutions out of the pool of possible solutions. Then their fitness is compared, and the better solution will be replicated. Hence, the tournament selection chooses the best individual in each process. This approach is capable at looking at parallel architecture (Fig. \ref{fig:tournament}).
 \begin{figure}[h]
 \centering
 \caption{Selection methods \cite{Parvez2014selection}}
 \begin{subfigure}{0.5\textwidth}
   \centering
   \includegraphics[width=.6\linewidth]{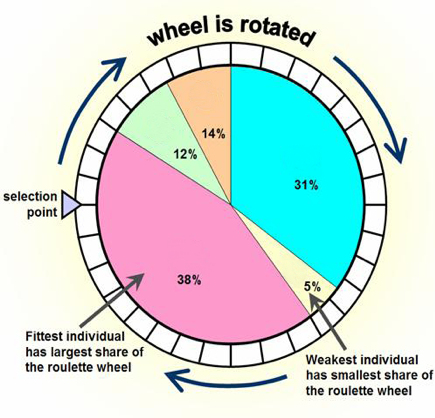}
   \caption{Roulette Wheel Selection}
   \label{fig:roulette}
 \end{subfigure}%
 \begin{subfigure}{.5\textwidth}
   \centering
   \includegraphics[width=.8\linewidth]{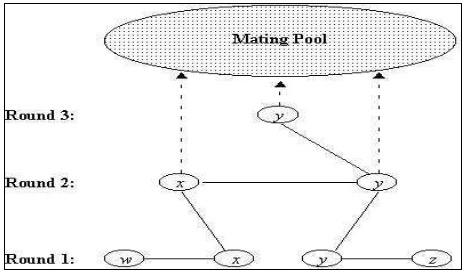}
   \caption{Tournament Selection}
   \label{fig:tournament}
 \end{subfigure}
 \label{fig:selection}
 \end{figure}
\item \textbf{Crossover: }
This genetic operator takes more than one parent solution and generates a child solution. Genes from parent chromosomes are taken and a new offspring is produced.
In detail, the operator selects a random crossover point ($\rvert$). Everything in the binary string from the first chromosome before this point is copied and everything after this point from the second chromosome is copied to generate a new offspring (Tab. \ref{tab:crossover}). 

\begin{table}[h]
 \caption{Single point crossover operator}
 \label{tab:crossover}
 \centering
 \begin{tabular}{| l | r |}
	 \hline
	 \colorbox{pink}{chromosome 1} & \colorbox{pink}{10100} $\rvert$ \colorbox{pink}{10100100110}\\
	 \colorbox{applegreen}{chromosome 2} & 
	 \colorbox{applegreen}{10100} $\rvert$ \colorbox{applegreen}{11000101100}\\
	 offspring 1 & \colorbox{pink}{10100} $\rvert$ \colorbox{applegreen}{11000101100}\\
	 offspring 2 & \colorbox{applegreen}{10100} $\rvert$ \colorbox{pink}{10100100110}\\
	 \hline
	 \end{tabular}
 \end{table}

It is possible to take several crossover points and ameliorate the performance of the GA by applying specific crossover on specific problems.
Crossover also permits the exchange of information in solutions, similar to natural organism reproduction \citep{coley1999introduction}.

\item \textbf{Mutation: }
This operator changes a single bit randomly in the current generated offspring from the crossover operator, e.g. from 0 to 1 or from 1 to 0 (Tab. \ref{tab:mutation}). 
The engine needs a way to produce a new guess by mutation the current one. The parent string is converted into an array with of parent strings, list(parent). After replacing 1 letter in this array with a random selected letter from the gene set, the result is recombined into a new string.
%
 \begin{table}[h]
 \caption{Mutation operator}
 \label{tab:mutation}
 \centering
 \begin{tabular}{| l | r |}
	 \hline
	 offspring 1 & 101\colorbox{pink}{0}0 11000101100\\
	 offspring 2 & 10100 1\colorbox{pink}{0}1001001\colorbox{pink}{1}0\\
	 mutated offspring 1 & 101\colorbox{applegreen}{1}0 11000101100\\
	 mutated offspring 2 & 10100 1\colorbox{applegreen}{1}1001001\colorbox{applegreen}{0}0\\
	 \hline
	 \end{tabular}
 \end{table}
The reason for mutation is to prevent that not all solutions of the population occur in the local optimum of the solved problem.

\item \textbf{Termination: }
The entire process is iterative until 
either the known solution has been found, the population of n-iterations has not changed, or a certain amount of time and generations have passed \cite{bajpai2010genetic}.
\end{enumerate}





\newpage
\section{Genetic Algorithms in Lens System Design} 
%
%
\subsection{First Studies of GA in Lens Design}
In 1990, research into the application of EAs for the monochromatic quartet was proposed at the International Lens Design Conference, now IODC \cite{o1991monochromatic}. Constraints of the four-element lens design included only spherical surfaces, Gradient-index (GRIN) optics elements, and Fresnel lenses.

Betensky successfully applied a GA to a Gaussian optics system design to correct aberrations, but mainly astigmatism. This algorithm was set with almost zero power operators including optimized structural changes in order to develop a lens system. 
To apply a GA to the zoom lens problem is arduous due to the complex requirements for each operator \cite{betensky1993postmodern}.
A few years later van Leijenhorst et al. investigated the GA as a tool for the automatic optimization and design of an optical lens system. Results showed fast and simultaneous corrected aberrations, even on less powerful computers. Further, they said it was possible that optical industries could use the GA for larger and more sophisticated optical systems \cite{van1996optical}.


Chen and Yamamoto applied a GA and a combined algorithm, including a GA and the DLS method, to lens design in order to correct lens aberrations and determine differences between these two methods. They found out that the GA is a
useful algorithm for the global search and a beneficial starting point to perform the DLS method. Moreover, with the GA it is possible to avoid the local minima in gradient-based algorithms because the GA is based on a merit function of a population 
\cite{chen1997experiment}.

In the 2000s, Ono et al. \cite{ono2000optimal} proposed a new lens optimization method by including real-coded GAs to small and large-scale problems. They applied their technique successfully for global and multi-objective optimization. It worked very effective in finding good lens designs for both small and large-scale problems. It was possible to analyze the lens design problem by optimizing a single criterion and with the two criteria distortion and resolution. The proposed GA utilizes unimodal normal distribution crossover (UNDX) and minimal generation gap (MGG) for an optimal lens design.

In terms of practical use the proposed GA by Ono et al. could not be applied for several reasons. By changing the thickness and curvature the GAs are not able to improve chromatic aberration. Due to this disadvantage, the algorithms are restricted to a single wavelength when designing optical systems such as IR cameras. Furthermore, the proposed GAs are not included in commercial CAD programs, which are essential for lens designers to use and do not provide an user interface that is easy to handle. 
A newer study by Fontana et al. \cite{fontana2008computer} showed the possible implementation of a GA into a CAD tool they had developed for optimization and design.

Banerjee and Hazra studied the application of a GA on a structural lens design. They effectively accomplished the search for a global solution within utilitarian local optima. It is not possible to ascertain all values for the various parameters in a GA all at once. Alterations can probably affect the efficiency of the GA. The results could be used to improve the GA for optimization problems in real life lens design \cite{banerjee2001experiments}.

Gagn{\'e} et al. 
applied EAs to an optimization criteria with complex mechanical constraints. Their results showed that EAs are comparable to those obtained by humans \cite{gagne2008human}. 

Thibault et. al. \cite{gagne2008human} discovered that on a real-life imaging problem the EA lens showed better results by a factor of almost two and was four times more sensitive than the expert lens design for the 1990 monochromatic quartet \cite{o1991monochromatic}. Moreover, an EA can explore a lens system, which is similar to a design performed by experts. Thibault et al. did not incorporate sensitivity with regard to lens bending and lens shape
\cite{thibault2005evolutionary}.

Cuevas et al. applied the GA to solve the closed fringe demodulation problem in 2006. The population of chromosomes in a GA is linked to the coefficients of polynomials to calculate the local phase of an interferogram window. Specifically, they used the multi-polynomial fitting (MPF) because the phase field is estimated and interpolated to any resolution size to assess the polynomial coefficients. In a second application, a GA was used on the nesting problem of finite materials. The nesting problem is based on the optimization of items which are nested in a valid place with the least wasted material \cite{cuevas2006genetic}.

Chen et al. used the GA in order to optimize Fresnel lenses which were inserted in light-emitting diodes (LED) sources. It is not feasible to guarantee sufficient flux with a single LED, but with numerous LEDs it is possible. Chen et al. proposed an optimal design of a Fresnel lens with several LEDs to provide white light. The GA, including a fitness function with the appropriate illuminance and uniformity, was employed to search for the optimal groove angles of the Fresnel lens. They figured out that the generated optimal Fresnel lens showed more of an augmented light-guidance than with typical Fresnel lenses for a multiple-LED reading light system. The problem is the intrinsically restricted execution speed while performing the nonimaging optical tool for the optimal design procedure because it is not designed to create an optimal design with more numbers of source light rays \cite{chen2006better}. 

Similar results were achieved by Chen et al. with their hierarchical GA (HGA) in order to optimize a large scale Fresnel lens for a multiple LED reading light system \cite{chen2006hierarchical}.

In 2013, Chen et al. suggested another method to optimize the optical LED design. Taguchi method \cite{taguchi1992taguchi} is well-functioning in the discontinuous region but poor for the use of an optimal optical design because of immoderate time consumption during the analyses. Angle and average illuminance were observed and indicated that the evolved system is applicable to the optical design of different LED lenses \cite{chen2013optimization}.

Fang et al. used a GA to effectively eliminate both axial and lateral chromatic aberrations of two Gauss-type lens designs and to find the appropriate glass combination of two different Gauss-type lenses. These two Gauss lens designs can remove primary chromatic aberrations. A telephoto lens design was used due to its sensitivity to axial aberration, in addition to a wide-angle Gauss design was used to deal with lateral chromatic aberrations. In comparison to the DLS method, the proposed method  to search for a two lens type combination was more successful \cite{fang2007eliminating}.

The lens designer has to deal with obstacles such as the variable nature of chromatic aberrations and influence. Fang et al. used the GA for lens design and optimization of the liquid lens element. The GA could be efficiently applied to replace the conventional least damping square (LDS) method to search for the optimal solution, including the chosen optical lens element but also to quickly select the best glass combination. They achieved promising results in searching for the nest optical layout with liquid lenses and the best glass set to remove chromatic aberrations. This research took into account only the first-order aberrations of thin lenses. Hence, the results were satisfactory
\cite{fang2008miniature}.

Other research by Fang et al. was focused on a new digital zoom layout and optimization using an altered GA. They successfully developed an optical design and optimization of digital zoom optics which includes the liquid aspheric lens surface and improved significantly the performance of zoom optics. They noted that the MTF resembles the conventional DLS method in terms of approaching the optical diffraction limit \cite{fang2011study}.

\subsection{State of the Art Technologies}
2010, Bajpai et al. applied the GA to achieve global optimization by using the Rastrigin’s function \cite{rastrigin1974extremal}. They determined with the Rastrigin's function many local minima function and only one global minimum. Further, they investigated the reasons why a GA is a good optimization tool. One of the main benefits is its intrinsic parallelity, and therefore the ability to assess various schemas at once
\cite{bajpai2010genetic}.

Furthermore, a hybrid GA is more effective and efficient than a traditional GA and is achieved if the GA includes other techniques in its framework. These kind of algorithms are inserted to solve real-world problems rapidly, precisely, and reliably, without any manual help \cite{bajpai2010genetic}.

The application of hybrid GAs in lens system design was studied by several authors \cite{fang2009extended,tsai2015improvement,yen2015aspherical1}.
Fang et al. proposed a hybrid Taguchi-GA to eliminate chromatic aberrations more effectively than with the conventional DLS method. This algorithm was applied to zoom optics with a diffractive optical element (DOE). Compared with the DLS method they found that chromatic aberration for optical lenses could be significantly reduced with the hybrid technology \cite{fang2009extended}. Also, Tsai et al. successfully applied the hybrid GA to zoom optics to specify the best position for DOE and to remove chromatic aberrations of the zoom optics with a DOE \cite{tsai2008extended}.

In 2015, Tsai et al. investigated the effect of a GA with DLS optimization on a projector lens to ameliorate the filed curvature aberration (FCA) and image resolution. Often optical software is not able to simultaneously optimize the FCA and image resolution of lens design. The software works either for global or local optimization and can only give out results of local optimization. The combination of a hybrid GA and DLS could be an optimal optimization approach for commercial software and has the capability to improve both image quality and various aberrations \cite{tsai2015improvement}.

Yen and Jin applied efficiently a GA on aspherical lens design to reduce aberrations in multifocal artificial intraocular lenses (IOL) \cite{yen2015aspherical2,yen2015aspherical1}. 
They inserted a GA by mimicking the variation of thickness and curvature of the human eye into the optical software CODE V to create an IOL design. By comparing the built-in software algorithm with the GA, the suggested GA for IOL design showed more improvements of the spot size in root mean square (RMS), tangential coma (TCO), and the MTF. 





\section{Conclusion} 

In this paper we have reviewed various results on the use of GAs in optical system design. A GA is applied to lens design to find the best global and stable solution within an optimization problem. Optimization is needed to improve the performance of a design. Compared to conventional optimization strategies (e.g. DLS), the GA is a powerful tool because it uses a coded parameter set, searches a population of points, and applies objective function information and probabilistic transition rules \cite{goldberg1989genetic}. Specific studies have proven that GAs are perfect for finding the optimal groove angles of the Fresnel lens, and are useful in finding the proper glass combination of two different Gauss-type lens designs. Several cutting edge studies successfully show that a hybrid-GA can be useful in lens design to correct and eliminate chromatic aberrations in the axial and lateral view. Fast and simultaneous corrections were applied to zoom optics with certain DOEs. The missing availability of the GA in commercial CAD programs has been resolved to a great extent.

However, there still exist some limitations on the use of GAs. Limitations of GAs include time consumption for convergence which can be improved by a proper sized population and many generations. Trial and error is connected with the fine tuning of all indispensable parameters for a GA. Defining the right definition of the MF for a GA is another crucial issue in lens design. Moreover, an inappropriate fitness function design may lead to unintelligible solutions. Having said this there is considerable scope for further research and exploration of GAs in optical engineering.


\newpage
\begin{appendices}

\section{Advantages \& disadvantages over other optimization strategies}
\label{appendix:adv}
\begin{center}
\vspace{-\baselineskip}
\linespread{1.0}
%
{\small \noindent
\begin{longtable}[h]{|p{7.0cm}|p{7.0cm}|}
\caption{Advantages \& disadvantages}\\ \hline
\rowcolor[gray]{0.9} 
\centering\textbf{Advantages} & \centering\textbf{Disadvantages} 
\endfirsthead
\multicolumn{2}{c}
{\tablename\ \thetable\ -- \textit{Continued from previous page}}\\ \hline 
\rowcolor[gray]{0.9}  
\centering\textbf{advantages} & \centering\textbf{disadvantages} 
\endhead \hline
\multicolumn{2}{r}{\textit{Continued on next page}}
\endfoot \hline
\endlastfoot
\label{tab:a-d}
%
%
	\tabitem Optimization with a systematized set of continuous or discrete parameters for global optimization scenarios.
	  & \tabitem Due to its stochastic nature, no convergence provided to find a global maxima.\\
     
      \tabitem No calculus of derivatives demanded.
      & \tabitem Instead of using only a single search value, it works with a population of solutions.\\
           
      \tabitem Works with a large number of parameters. 
      & \tabitem High computational cost is required to look into billions of solutions which impede the optimization of hard problems \cite{thibault2005evolutionary}.\\ 
   
      \tabitem Intrinsically parallelity permits the analysis of many schemas at once.
      & \tabitem Goldbergs Pascal code of a simple GA from 1989 
      was used for more than 25 years but can be seen as obsolete and useless for nearly all real-value decision.\\   
    
      \tabitem Optimization of complex objective functions and criteria \cite{bajpai2010genetic,cuevas2006genetic}.
      & \tabitem The binary representation is not appropriate for real-valued-decisions \cite{goldberg1989genetic}.
      \\ 
      
      \tabitem Easy to use for black-box simulation modelling \cite{paul1998simulation}. 
      & \tabitem Convex optimization techniques presume a functional relationship between decisions and objectives \cite{paul1998simulation}.\\ 

      \tabitem Overcome border of local optimums.
      & \tabitem Time consumption for convergence.\\
  
      \tabitem The result is a set of solutions and not a single solution.
      & \tabitem Fine tuning of all parameters for the GA is associated with trials and errors.\\
   
      \tabitem Able to work with experimental data or numerically generated data \cite{cuevas2006genetic}.
      & \tabitem Efficacy of the system depends strictly on the fitness function, choice of genetic encoding, and genotype to phenotype mapping \cite{whigham2017mapping}.\\ 
 
      \tabitem Optimization over a broad search space is attainable due to an increased population of chromosomes \cite{chen1997experiment}.
     & \tabitem The fitness function needs to go through dynamic scaling before selection \cite{coley1999introduction}.\\ 
 
     \tabitem Efficient and comprehensive search method for optimization.
     & \tabitem Simulated annealing and deterministic global searching \cite{bajpai2010genetic}.\\
 
    \tabitem Characterizable and controllable procedure of alteration. 
  	& \\
    
    \tabitem Multiple offspring and are able to examine the solution space in multiple directions. 
	& \\
    
    \tabitem Well-suited for a fitness function which is discontinuous, noisy and changes over time, or has several local optima. 
    & \\
    
     \tabitem GAs can manipulate numerous parameters simultaneously. The parallelism allows them to generate different equally good solutions to an identical problems.
     & \\
     
     \tabitem GAs are unaware 
     of the problems in which they are applied, because of random changes within candidate solutions.
     \cite{bajpai2010genetic}. 
     & 
\label{table:adv}
\end{longtable}
}

\end{center}

\newpage
\section{Historical Perspective}
\label{appendix:historical}
For more than 20 years GAs have been successfully applied in the areas of image processing \cite{chen1997blind,kawaguchi1997}, medicine \cite{yang1998new}, or laser technology \cite{carroll1996chemical,carroll1996genetic}.

Table \ref{table:historical} gives the historical perspective on the use of GA in lens design.

\vspace{-0.4cm}
\begin{center} \footnotesize
\linespread{1.0}

\begin{longtable}[h]
{|>{\columncolor[gray]{0.9}}p{0.7cm}|p{8.3cm}|p{3.0cm}|p{1.08cm}|}
\caption{Historical perspective on selected studies of GAs for lens design}\\ \hline 
\rowcolor[gray]{0.9} 
\textbf{year} & \textbf{study} & \textbf{author} & \textbf{ref.}\\ \hline
\endfirsthead

\multicolumn{4}{c}
{\tablename\ \thetable\ -- \textit{Continued from previous page}}\\ \hline 
\rowcolor[gray]{0.9}  
\textbf{year} & \textbf{study} & \textbf{author} & \textbf{ref.}\\ \hline
\endhead \hline
\multicolumn{4}{r}{\textit{Continued on next page}}\\
\endfoot \hline
\endlastfoot

	  1963 & \tabitem First experiments in ESs & Zbigniew M. & \cite{michalewicz1996evolution,schwefel1981numerical}\\
      
      1973 & \tabitem First publication of strategy of evolution & Rechenberg I. & \cite{rechenberg1973evolutionsstrategie}\\
      
      1975 & \tabitem Strategy of evolution and first GA developed & Holland J. & \cite{holland1975adaptation}\\
      
      1988 & \tabitem Overview of GAs & De Jong & \cite{de1988learning}\\
      
      1993 & \tabitem A GA is a robust method to evaluate unknown parameters in a physical system & De Jong & \cite{de1993genetic}\\
      & \tabitem GA of structural changes on Gaussian optics system design & Betensky & \cite{betensky1993postmodern}\\
      
      1996 & \tabitem An optical design using a GA & Van Leijenhorst et al. & \cite{van1996optical}\\
      & \tabitem Optimization of lens design using a GA & V\'{a}zquez-Montiel and Cornejo-Rodr\'{i}guez & \cite{vazquez1996lens}\\
       & \tabitem Investigation of GAs to optimize lens design & Chen and Yamamoto & \cite{chen1996genetic}\\
                  
      1999 & \tabitem Real-coded GAs with UNDX to design a lens system which includes glass selection & Ono et al. & \citep{ono1999designing}\\
      
      2000 & \tabitem Development of real-coded GAs using UNDX and MGG for an optimal lens design & Ono et al. & \cite{ono2000optimal}\\
      
      2001 & \tabitem Proposals for experiments with GAs for structural design of cemented doublets & Banerjee and Hazra & \cite{banerjee2001experiments}\\
      
      2002 & \tabitem GAs on lens system design and re-engineering experiments & Beaulieu et al. & \cite{beaulieu2002lens}\\
      
      2003 & \tabitem Introduction of GA for lens design with cemented triplets which is a thin lens model using curvatures of four interfaces & Chatterjee et al. & \cite{chatterjee2004structural}\\
      
      2004 & \tabitem Optical device design using GAs & Sanchis et al. & \cite{sanchis2004integrated}\\
            
      2006 & \tabitem GA in the field of \textit{closed fringe demodulation} in optics and engineering and used multi-polynomial fitting (MPF) & Cuevas et al. & \cite{cuevas2006genetic}\\
      & \tabitem GA used to optimize Fresnel lens for a multiple-LED reading light system & Chen et al. & \cite{chen2006better}\\
      & \tabitem A HGA-based approach applied to the optimal design of the Fresnel lens for a reading light system that consists of multiple LED sources & Chen et al. & \cite{chen2006hierarchical}\\
      
      2007 & \tabitem Method to eliminate chromatic aberrations in Gauss-type lens design with the use of a novel genetic algorithm & Fang et al. & \cite{fang2007eliminating}\\

      2008 & \tabitem Lens design and optimization method with GA applied to liquid lens elements & Fang et al. & \cite{fang2008miniature}\\
      & \tabitem Hybrid GA to optimize a 350X zoom optics system  & Tsai et al. & \cite{tsai2008extended}\\
      & \tabitem ES applied to a human-competitive lens system design & Gagn{\'e} et al. & \cite{gagne2008human}\\
      & \tabitem Application of evolutionary algorithms to lens design & Thibault et al. & \cite{thibault2005evolutionary}\\
    
    2009 & \tabitem Hybrid Taguchi-GA applied to zoom optics including a DOE & Fang et al. & \cite{fang2009extended}\\
    
      2011 & \tabitem An altered GA applied to optical design and optimization of the digital zoom optics with liquid lenses & Fang et al. & \cite{fang2011study}\\
      
      2013 & \tabitem Systematic method used to optimize an optical design for a LED lens module & Chen et al. & \cite{chen2013optimization}\\
      
      2014 & \tabitem GAs used to optimize optical systems & Neda HesamMahmoudiNezhad & \cite{hesammahmoudinezhad2014optical}\\
            
      2015 & \tabitem Investigation of the influence of aberrations on contact lenses and a hybrid neural-GA was used to optimize the aspherical lens design of myopic and astigmatic eyes
      & Yen and Ye & \cite{yen2015aspherical1}\\
      & \tabitem GA used to design a multifocal IOL design for the human eye & Yen and Jin & \cite{yen2015aspherical2}\\
      & \tabitem Hybrid genetic algorithm used to improve field curvature aberration in optical lenses & Tsai and Fang & \cite{tsai2015improvement} 
\label{table:historical}
\end{longtable}
\end{center}


\end{appendices}

\newpage
\bibliography{files/bibfile}   
\bibliographystyle{spiejour}   

\newpage



\end{spacing}
\end{document}